\documentclass{emulateapj}
\usepackage{apjfonts}

\begin{document}

\title{Evidence for a Stellar Disruption by an IMBH
in an Extragalactic Globular Cluster\altaffilmark{1}}
\shorttitle{Stellar Disruption by a Globular Cluster IMBH}

\author{Jimmy A. Irwin\altaffilmark{2,3}, Thomas G. Brink\altaffilmark{3},
Joel N. Bregman\altaffilmark{3} and Timothy P. Roberts\altaffilmark{4}}
\email{jairwin@ua.edu}
\shortauthors{Irwin et al.}

\altaffiltext{1}{This paper includes data gathered with the 6.5 meter Magellan
Telescopes located at Las Campanas Observatory, Chile.}
\altaffiltext{2}{Department of Physics and Astronomy, University of Alabama,
Box 870324, Tuscaloosa, AL 35487}
\altaffiltext{3}{Department of Astronomy, University of Michigan,
500 Church St., Ann Arbor, MI 48109-1042}
\altaffiltext{4}{Department of Physics, Durham University, South Road,
Durham DH1 3LE, UK}

\begin{abstract}
We report [O III] $\lambda$5007 \AA\ and [N II] $\lambda$6583 \AA\ emission
from a globular cluster harboring the ultraluminous X-ray source (ULX)
CXOJ033831.8-352604 in the Fornax elliptical galaxy NGC~1399. No 
accompanying Balmer emission lines are present in the spectrum.
One possibility is that the forbidden lines emanate
from X-ray illuminated debris of a star that has been tidally-disrupted by
an intermediate-mass black hole (IMBH), with this debris also feeding the black
hole leading to the observed X-ray emission. The line strengths indicate that
the minimum size of the emitting region is $\sim$$10^{15}$ cm, and if the
70 km s$^{-1}$ half-widths of the emission lines represent rotation around
the black hole, a minimum black hole mass of 1000 M$_{\odot}$ is implied.
The non-detection of H$\alpha$ and H$\beta$ emission lines suggests a white
dwarf star was disrupted, although the presence of strong nitrogen emission
is somewhat of a mystery.
% We compare CXOJ033831.8-352604 to another ULX
%in the extragalactic globular cluster RZ2109 that also harbors strong [O III]
%and lacks Balmer lines. While the $\sim$1500 km s$^{-1}$ wide [O III] emission
%of RZ2109 has been cited as evidence for a wind from a stellar-mass black hole,
%we suggest an alternative scenario where the wide [O III] line results from the
%aftermath of a supernova event
%induced by the tidal-detonation of a white dwarf by an IMBH. Further
%modeling of such an event is needed to determine if this model explains
%the broad [O III] line as well as the stellar-mass black hole wind model.

\end{abstract}


\section{Introduction}

Despite decades of effort, finding clear-cut evidence for the existence of
intermediate-mass black holes (IMBHs; $100-10^4$ M$_{\odot}$) within 
globular clusters has proved quite
elusive.
Initial claims of kinematical evidence for a $>$1000 M$_{\odot}$
black hole in the Milky Way globular
M15 as far back as the 1970s (Newell, Da Costa, \& Norris 1976)
have been vigorously questioned over the years (Illingworth \& King 1977;
McNamara et al.\ 2003). Further claims of IMBHs in the Milky Way globular
clusters NGC~6388 (Lanzoni et al.\ 2007) and Omega Cen (Noyola, Gebhardt, \&
Bergmann 2008) as well as G1 in M31
(Gebhardt, Rich, \& Ho 2002) also have not been met with universal
acceptance (Baumgardt et al.\ 2003; Nucita et al. 2008;
van der Marel \& Anderson 2009).

%Along a different line of reasoning, ultraluminous X-ray sources (ULXs)
%have been suggested to be powered by accretion onto
%an IMBH by a stellar companion. The X-rays luminosities of the brightest ULXs
%exceed the Eddington limit of a 10 M$_{\odot}$ black hole by over an order of
%magnitude (Colbert \& Mushotzky 1999). In addition, the measured apparent
%temperatures of the inner accretion disks of some ULXs (expected to scale as
%$m^{-0.25}$) are considerably lower than in stellar-mass Galactic black hole
%binaries (e.g., Miller et al.\ 2003), suggesting a higher mass.
%However, alternative
%explanations for the observed properties of ULXs that do not invoke an IMBH
%are equally strong (Roberts 2007; King 2008).

If IMBHs exist within globular clusters
they should occasionally disrupt passing stars, either tidally-shredding
them or inducing a supernova explosion in the event a white dwarf (WD) strays
too close (Frank \& Rees 1976; Baumgardt, Makino, \& Ebisuzaki 2004;
Rosswog, Ramirez-Ruiz, \& Hix 2008; Ramirez-Ruiz \& Rosswog 2009).
Such an event would leave unique transitory optical, ultraviolet, and X-ray
signatures lasting years to perhaps centuries.
In this {\it Letter}, we present the discovery of [N II] $\lambda$6583 \AA\ and
[O III] $\lambda$5007 \AA\ emission from the globular cluster that
harbors the ULX CXOJ033831.8-352604 in the Fornax elliptical galaxy NGC~1399.
We explore the idea that the X-ray emission and optical emission lines
result from the aftermath of the tidal disruption of a WD by an IMBH
in this globular cluster.
%We also explore the idea that another optical forbidden line-harboring ULX in
%the globular cluster RZ2109 within the Virgo elliptical
%galaxy NGC~4472 discussed in Zepf et al.\ (2007, 2008)
%might be explained by the aftermath of a supernova event induced by the
%tidal detonation of a WD by an IMBH.
%We also investigate explanations that accommodate both the X-ray and optical
%properties of the two sources in terms of stellar-mass black holes, and suggest 
%how future observations can distinguish between the two mass classes.

\section{Optical and X-ray Observations}

CXOJ033831.8-352604 was first identified as a globular cluster X-ray source
by Angelini et al.\ (2001), with the host globular cluster being
quite luminous and red ($B-I=2.25$, $M_I=-10.5$;
Kundu, Maccarone, \& Zepf 2007 assuming $d = 20$ Mpc).
We observed the globular cluster for 1 hour on 2005
October 10 with the Low Dispersion Survey Spectrograph (LDSS-3) mounted
on the Magellan II Clay 6.5-m Telescope (Las Campanas Observatory, Chile),
again on 2006 November 26--27 for 4 hours
with the Inamori Magellan Areal Camera and Spectrograph (IMACS) on the
Magellan I Baade Telescope, and finally on 2008 October 27 for 7.5 hour with
the Magellan Echellette Spectrograph (MagE) on the Magellan II Clay Telescope.
Superimposed on the continuum emission from the globular cluster is
an emission line at 6613 \AA\ detected in all three observations, and a
second emission line at 5029 \AA\ detected in the final two observations
(the LDSS-3 observation was not sensitive below 6000 \AA). Both emission lines
are doublets, and correspond to [N II] $\lambda$6583 \AA\ and [O III]
$\lambda$5007 \AA\ at a recessional velocity of 1360 km s$^{-1}$. This
is consistent with the velocity measured for the Ca II triplet
(1360--1372 km s$^{-1}$) and Na I D doublet (1354--1362 km s$^{-1}$)
absorption lines in the globular cluster spectrum
and with the recessional velocity of NGC~1399 itself.

The emission lines were resolved in the MagE spectrum (instrumental
FWHM = 55 km s$^{-1}$) with a FWHM of 140 km s$^{-1}$. The luminosity
of each line was $few \times 10^{36}$ ergs s$^{-1}$.
The regions of the flux-calibrated MagE spectrum around the two emission
lines are shown in Figure~1. Figure~1
also illustrates the lack of H$\alpha$ in the spectrum.
In determining the uncertainties to line emission, we made fits to the
emission lines, using a Gaussian line profile, where the fitted
quantities were the continuum level, line center, line width, and line
amplitude.  Fits to the lines had acceptable $\chi^2$ values and errors
were calculated in the usual way, as $\chi^2$ deviations that correspond to
1$\sigma$, etc.  Upper limits to undetected line emission were obtained by
similar line fitting techniques, except that the line center was fixed (so we
assumed that all lines have the same velocity) and the line width was
fixed at the value from other lines.
A line amplitude and its error
was determined in each case.  The line amplitude had a S/N $<$ 2 in those
cases (often, less than 1$\sigma$ and often with the wrong sign).  Based
on the error, 3$\sigma$
upper limits were formed. Balmer absorption from the globular cluster
could be affecting the detection of any Balmer emission lines. Using the
spectra of other globular clusters of similar luminosity, metallicity, and S/N
obtained with the same IMACS setup, we did not detect any Balmer absorption
lines, and found the upper limits on
Balmer features were no different than in CXOJ033831.8-352604. This indicates
that statistical uncertainties in the continuum level dominate any Balmer
absorption. Therefore,
we use the IMAC-only determination of the upper limit on Balmer emission in
our line ratio calculations.
By summing the
statistics of all three observations for [N II] and [O III], we can place a 3$\sigma$ lower limit on
[N II]/H$\alpha$ of 7. We can also place a 3$\sigma$ lower limit on
[O III]/H$\beta$ of 5.
% This calculation neglects the effect of fill-in
%of the Balmer absorption lines by emission at these wavelengths.
%We note that line fill-in of the H$\alpha$ absorption
%line is probably not significantly affecting this result, since the width of
%the absorption line (set by the 1.3 \AA\ FWHM resolution of the MagE spectrum)
%would be considerably narrower than the 3.1 \AA\ FWHM of the expected
%emission line, which would lead to an emission line with an absorbed core. Such
%a feature is not apparent in the spectrum.
Utilizing the IRAF line diagnostic routine {\sc temden},
we can use the lack of a [N II] $\lambda$5755 \AA\
detection to place a 3$\sigma$ upper limit on the temperature of the emitting
region of 13,000 K if a single-temperature model with a density equal to
the critical density of [N II] is assumed.

To determine whether we should have detected the double-horn profile indicative
of rotation, we convolved a double-horn
velocity profile with a Gaussian of 55 km/s (the spectral resolution of MagE).
For the rotational line shape, we took a rotating ring, which produces the
sharpest double-horned profile.  The rotational velocity was chosen so 
that the convolved profile has the same width as the observed profile.  
At our spectral resolution, the double-horned profile appears nearly as 
a single-peaked profile with only a small local minimum at line center.  
At the S/N of our data, this model is not distinguishable from a Gaussian
profile.
%In the future, a higher spectral resolution observation can 
%investigate this further.

CXOJ033831.8-352604 has been observed with {\it Chandra} on eight occasions
over the past decade.
The data were processed in a uniform manner
following the {\it Chandra} data reduction threads using {\sc ciao}v3.4.
An extraction region representing the 90\% enclosed flux
aperture for a point source at the position of CXOJ033831.8-352604 on the
detector was assumed. Spectra were extracted and responses were generated
using the {\sc ciao} tool {\sc specextract} with local background chosen from
an annulus surrounding the source. The spectra were
grouped to contain 25 counts per channel. Channels with energies
$<$0.5 keV and $>$8.0 keV were excluded. The spectra were fit
within {\sc xspec} with a simple power law model absorbed by the Galactic
hydrogen column density toward NGC~1399 ($N_H = 1.34 \times 10^{20}$
cm$^{-2}$; Dickey \& Lockman 1990). A disk blackbody model absorbed
by the Galactic hydrogen column density was also employed.

Only three of the {\it Chandra} observations yielded a
useful spectrum ($>100$ counts). Observations 0319 (2000 January 18),
4172 (2003 May 26), and 9530 (2008 June 6)
yielded best-fit power law exponents with 90\% uncertainties
of $2.5 \pm 0.2$, $3.0 \pm 0.5$, and
$2.5 \pm 0.4$, and 0.3--10 keV luminosities of $2.3, 1.6$, and $1.5 \times
10^{39}$ ergs s$^{-1}$, respectively, with $\chi_{\nu}^2$ values near unity.
Using a disk blackbody model instead yielded best-fit temperatures of
$0.38 \pm 0.04$, $0.39 \pm 0.08$, and $0.36 \pm 0.07$ keV, respectively, and
luminosities about 30\% lower than in the power law fits.
%The five shorter {\it Chandra} observations yielded luminosities consistent
%with the three longer observations.

\section{Unlikeliness of Planetary Nebula or Supernova Remnant Explanations}
Optical emission lines emanating from extragalactic globular clusters are
unusual, with an estimated occurrence rate of a few percent at most
based on literature searches. Those found are usually attributed to
planetary nebulae (Larsen 2008) or supernova remnants
(Peng, Ford, \& Freeman 2004).
% or from an unknown source (Pierce et al.\ 2006).
Chomiuk, Strader, \& Brodie (2008) found two [O III]
emission-line globular clusters in the S0 galaxy NGC7~457 which are believed
to be a planetary nebula and a supernova remnant. However, none of these
globular clusters are associated with luminous X-ray sources.
Globular clusters harboring X-ray sources
with $L_X > 2 \times 10^{39}$ ergs
s$^{-1}$ are even rarer; CXOJ033831.8-352604 is the third brightest X-ray
source in a globular cluster of the tens of thousands of globular clusters
around all galaxies imaged by {\it Chandra} within 20 Mpc.
The odds of both events occurring independently in
an average globular cluster are $\la10^{-5}$, so the odds for a ten times
more massive globular cluster like the one harboring CXOJ033831.8-352604
is $\la10^{-3}$. We will therefore focus on
explanations that account for both the high X-ray luminosity and optical
emission lines in terms of a single object.

While planetary nebulae are sources of strong [O III]/[N II] emission,
their X-ray luminosities are orders of magnitude lower
than CXOJ033831.8-352604. Furthermore, very few planetary nebulae in the Milky
Way have [N II]/H$\alpha$ ratios as high as 7 (Riesgo \& L\'opez 2006), nor do
they have expansion velocities of 70 km s$^{-1}$.
Supernova remnants can be strong [O III] emitters and on rare
occasion have $L_X > 10^{38}$ ergs s$^{-1}$ (Patnaude \& Fesen 2003;
Bauer et al. 2008).
However, in order for the expansion of
the remnant to have slowed down from an initial velocity of $\sim$10$^4$
km s$^{-1}$ to the measured velocity of $\sim$70 km s$^{-1}$ (assuming the width
of the line is interpreted as expansion rather than rotation), the remnant
would have needed to sweep up a substantial amount of ISM
within the globular cluster. Assuming there is even enough ISM
(primarily hydrogen) in the globular cluster to slow down the expansion,
we would expect to detect significant H$\alpha$ emission as is seen in
most supernova remnants. Furthermore, supernova remnants do not show
such large [N II]/H$\alpha$ ratios (Payne et al.\ 2007;
Payne, White, \& Filipovi\'c 2008).

%The luminous, persistent nature of the X-ray source
%argues that an
%accreting black hole is the source of the X-ray emission. For more than
%eight years, the source has shone at ten times the Eddington limit of
%a neutron star.
%While a growing number of ULXs
%located within or near star-forming regions of spiral galaxies are found to
%harbor optical emission lines (Pakull \& Mironi 2002; Abolmasov et al.\ 2007;
%Kaaret \& Corbel 2009), they are unlikely to represent
%the same phenomenon seen in CXOJ033831.8-352604. The emission
%lines in some spiral galaxy ULXs are associated with resolved nebula
%that appear to be bubbles blown in the surrounding dense interstellar medium by
%the ULX jet/stellar wind, although some are associated with bright central
%X-ray ionized regions. H$\alpha$ is also detected with fluxes comparable to
%or exceeding the [N II] flux. For CXOJ033831.8-352604, the lack of any
%substantial dense interstellar medium expected in a globular cluster
%within an elliptical
%galaxy coupled with the very unusual [N II]/H$\alpha$ ratio
%indicates the source of the emission line gas must be very local to the
%black hole and furthermore, must emanate from material severely lacking in
%hydrogen.

\section{Stellar Disruption by an IMBH?}
The luminous, persistent nature of the X-ray source
argues that an
accreting black hole is the source of the X-ray emission. For more than
eight years, the source has shone at ten times the Eddington limit of
a neutron star.
The X-ray luminosity of CXOJ033831.8-352604 is consistent with either a
stellar-mass black hole accreting near its Eddington limit, or an IMBH
accreting at 0.1--1\% of its Eddington limit. 
%While we discuss stellar-mass black holes below,
Here, we explore the idea that the observed
X-ray and optical properties of CXOJ033831.8-352604 are best-described
as the aftermath of a tidal disruption event of a star passing by an
IMBH within the globular cluster.

The idea that IMBHs within globular clusters will tidally
disrupt passing stars has been increasingly explored recently.
N-body simulations show a 1000 M$_{\odot}$
IMBH at the center of a globular cluster should tidally disrupt a passing star
every $10^5-10^9$ yr depending on the central density of the globular cluster
(Baumgardt et al.\ 2004). A host of studies
(involving both $10^6$ M$_{\odot}$ supermassive black holes in galactic nuclei
and IMBHs in globular clusters) predict the debris from the disrupted
star forms a precessing,
self-interacting stream, which ultimately forms an accretion disk, an
optically-thick envelope, and a quasi-spherical $\sim10^4$ K diffuse
photosphere extending
to $\sim$$10^{15}$ cm or larger. Accretion of the debris material
by the black hole leads to an intense, short-lived (few months)
UV/soft X-ray
flare peaking near the Eddington limit (Rees 1988; Loeb \& Ulmer 1997;
Ulmer et al.\ 1998) followed by a less intense accretion phase lasting
much longer (Cannizzo, Lee,
\& Goodman 1990; Ulmer et al.\ 1998) where the X-ray luminosity is predicted to
decline as $t^{-5/3}$. During this less intense accretion phase,
it is expected that the optically-thick envelope and diffuse photosphere from
the stellar debris reprocesses the X-ray emission from the accretion disk
to the optical/UV part of the spectrum
(Loeb \& Ulmer 1997; Bogdanovi\'c et al.\ 2004; Sesano, Erackeous, \&
Sigurdsson 2008).
Work by Ramirez-Ruiz \& Rosswog (2009) indicate that an IMBH-induced tidal
disruption event could continue to emit at $L_X > 10^{39}$ ergs s$^{-1}$ for
more than a century after the disruption event,
and suggest that line emission from globular clusters
might earmark the presence of recent tidal disruption events.
A variation on the tidal disruption theme is if the passing star is a white
dwarf that
experiences such strong tidal forces by an IMBH that its core is compressed
to the point thermonuclear detonation is triggered (Rosswog et al.\ 2008).
An underluminous, Type Ia supernova-like event would then occur with debris
being ejected from the globular cluster at high velocities.
% although likely
%at lower velocities than a traditional Type Ia supernova.

The modest widths of the [O III] and [N II] lines of CXOJ033831.8-352604
argues against a thermonuclear {\it detonation} tidal event, but might
represent rotation of material around the black hole following
a tidal {\it disruption} event. In this scenario, the X-ray
emission results from debris material falling onto the black hole, while
the [O III] and [N II] emission lines emanate from the reprocessing of
escaping X-ray photons by material in the diffuse photosphere.
The upper limit on the temperature of the
line-emitting region of 13,000 K from [N II] line diagnostics
is consistent with the predicted temperature of the diffuse photosphere.
The 35\% decline in X-ray luminosity from 2000 to 2008 is not necessarily
incompatible with the predicted $t^{-5/3}$ decline if the actual tidal
event took place a century ago. The soft X-ray spectrum is
also similar to suspected tidal disruption events from supermassive
black holes, such as in NGC~3599 and SDSS J132341.97+482701.3, whose X-ray
spectra remained soft ($\Gamma \sim 3$) several years after discovery,
when the X-ray emission had faded three orders of magnitudes from
their Eddington flare peaks (Esquej et al.\ 2008).

With the assumption the [O III] and [N II] lines represent rotation of debris
material around an IMBH, we can use 
the strength of the forbidden [O III] and [N II] lines to
place a lower limit on the size of the emitting region, and therefore a lower
limit on the mass of the IMBH. The luminosity of the line can be
expressed as L = $\epsilon_{\nu}$~$n_e$~$n_{ion}$~$V$, where $\epsilon_{\nu}$
is the emissivity of the line per unit electron ($n_e$) and unit ion
($n_{ion}$) densities, and the emitting
volume, $V$, is 4/3$\pi R^3$. We can place an upper limit on $n_e$ of
$8 \times 10^4$ cm$^{-3}$ since this is the critical density of [N II], and we
would not observe [N II] if $n_e$ were substantially higher. Assuming
$n_{ion} \sim n_e$, we can then place a lower limit on R, the physical size
of the emitting region. The [N II] line has a luminosity
of $3 \times 10^{36}$ ergs s$^{-1}$, leading to a lower limit on R of
$\sim$$3\times 10^{15}$ cm.
Coupled with a rotation velocity of 70 km s$^{-1}$,
this leads to a lower limit of the black hole mass of $\sim$1000 M${_\odot}$.
%Better constraints could be placed on the black hole mass if an
%estimate of the density of the emission line region could be obtained.
%C III] 1907 \AA\ /1909 \AA\ ratio is a sensitive indicator of
%gas density. A {\it Hubble Space Telescope Cosmic Spectrograph} ultraviolet
%spectrum of CXOJ033831.8-352604 would be useful to search for C III] emission.

The argument that we present does not critically depend on the gas being in 
rotational equilibrium.  It depends on the line width being indicative 
of the depth of the potential well.  For non-explosive events (not 
supernova or nova events), the outflow, inflow, or turbulent velocity is 
expected to have an energy similar to the potential well at that point, 
$v^2 \approx GM/R$.  This is the situation for the broad line region in 
AGNs, for example, which is used to determine masses of supermassive black
holes.  In the AGN case, the distance to the emitting line region comes from
reverberation line mapping.  In our case, the lower limit on the distance
is determined from the need for the emitting region to have a 
density below the critical density of [N II].

Another argument in favor of an IMBH rather than a stellar-mass black hole
for the disrupting object is the expected timescale of decline of the X-ray
emission following the disruption event. Following the Eddington flare,
the X-ray luminosity should decay as $t^{-5/3}$ (and possibly as $t^{-1.2}$
at later times; Ramirez-Ruiz \& Rosswog 2009).
A $<$100 $M_{\odot}$ black hole would have a peak X-ray luminosity of
a $<$$few \times 10^{40}$ ergs s$^{-1}$.
A 100 $M_{\odot}$ black hole would have to have
been accreting at 10\% its Eddington limit in 2000, indicating a disruption
event no more than a year earlier. By 2008, the X-ray emission should have
declined two orders of magnitude, rather than the measured 35\%.
However, a several thousand solar mass black hole would flare to
$\sim$$5 \times 10^{41}$ ergs s$^{-1}$, and would take some time (decades) to
decay down to the observed $1.5-2.3 \times 10^{39}$ ergs s$^{-1}$,
%(as suggested by Ramirez-Ruiz \& Rosswog 2009)
indicating the disruption took place some time ago.

The lack of Balmer emission in the spectrum suggests that the disrupted star
was a white dwarf. However, the presence of such a strong
nitrogen emission line from WD material is somewhat puzzling.
The AM Cvn star GP Com (a double WD binary) shows
nitrogen and oxygen in the optical (Marsh, Horne, \& Rosen 1991), and nitrogen
in X-ray (Strohmayer 2004), although the unusual evolutionary path needed to
create a system like GP Com seems untenable in a binary with an IMBH primary.
There are a few ultracompact binaries believed to
have helium WD star donors for which significant nitrogen is detected
from the accretion disk (Nelemans, Jonker \& Steeghs  2006),
but in these instances helium
is also clearly detected, unlike in CXOJ033831.8-352604. Further modeling
of the optical spectrum of CXOJ033831.8-352604 will be required to determine
if such large amounts of helium could be present but undetected given the
physical conditions of the gas that is emitting the [N II]/[O III] lines.
If the disrupted star was a helium WD, then our assumption that
$n_{ion} \sim n_e$ is inaccurate, in the sense that a lower $n_{ion}$ is
implied for a given $n_e$. This will increase the lower limit of the mass
estimate of the black hole.

%While unlikely, it is possible that the WD accreted N-rich
%material from a companion star previous to being tidally-disrupted, with the
%donor star eventually being replaced by the IMBH in a three-body encounter.

%\section{Comparison to RZ2109 in NGC~4472}
There is another source that shares similarities with
CXOJ033831.8-352604.
Zepf et al.\ (2007, 2008) reported [O III] $\lambda$5007 \AA\
emission from
the ULX-harboring globular cluster RZ2109 in the Virgo elliptical galaxy
NGC~4472. Both RZ2109 and CXOJ033831.8-352604
exhibit strong [O III] emission, both show a
lack of H$\alpha$ and H$\beta$ emission, and the X-ray spectra of both
sources are quite soft. However, while CXOJ033831.8-352604
harbors strong [N II] emission with a flux greater than the [O III],
the upper limit on the [N II]/[O III] ratio for RZ2109 is 
only a few percent. Also, the [O III] luminosity of RZ2109 is
an order of magnitude larger than in CXOJ033831.8-352604.
Furthermore, the measured width of the [O III] line
of RZ2109 is $\sim$1500 km s$^{-1}$, whereas the widths of the [N II]
and [O III] lines of CXOJ033831.8-352604 are an order of magnitude narrower.
%Finally, the globular cluster hosting
%CXOJ033831.8-352604 is red and quite metal-rich, while RZ2109 is metal-poor
%(Maccarone et al.\ 2007).

%The lack of hydrogen in both objects argues against ionization of interstellar
%medium within the globular cluster, and severely limits the choices for the
%source of the line-emitting material. It seems inescapable that the material
%must originate from a C/O white dwarf to account for the strong [O III]
%emission. A white dwarf explanation for the line-emitting
%material is further enforced by the vastly different
%[O III]/[N II] line ratio between the metal-rich globular cluster
%harboring CXOJ033831.8-352604 and the metal-poor globular cluster RZ2109.
%The nitrogen content of all stars of a globular cluster scales with the
%metallicity of the globular cluster, as does oxygen. This suggests 
%an object highly overabundant in oxygen relative to nitrogen
%must be responsible for the line
%emission from RZ2109. Such an overabundance in oxygen in a metal-poor
%environment is naturally explained by a C/O white dwarf.

Zepf et al.\ (2008) point out that the $\sim$1500 km s$^{-1}$
width of the [O III] line of
RZ2109 is too broad to represent rotation around a black hole unless
the black hole is unrealistically massive. Instead, Zepf
al.\ (2008) argue that the most reasonable explanation for the X-ray/optical
properties of RZ2109 is that the X-ray emission from the black hole is
photoionizing material blown by a black hole wind. They argue
that since AGN winds are only blown by black holes accreting near
their Eddington limit, the black hole in RZ2109 must also be accreting
near its Eddington limit. As the peak X-ray luminosity of RZ2109 is $4 \times
10^{39}$ ergs s$^{-1}$, a stellar-mass black hole is implied.
In this case, the donor star is most likely a WD to explain
the lack of hydrogen lines, making the
system an ultracompact binary.

%However, the optical spectrum of RZ2109 also bears a striking resemblance to
%the O-rich supernova remnant in NGC~4449 and 1E 0102.2-7219 in the SMC
%(Milisavljevic \& Fesen 2008; Blair et al.\ 2000). The supernova remnant in
%NGC~4449 exhibits multiple [O III] velocity peaks like RZ2109. Blair et al.\
%(2000) suggest the progenitor of 1E 0102.2-7219 was most likely a
%Wolf-Rayet star with a large oxygen mantle. While
%the object in RZ2109 cannot be a core-collapse supernova like the ones in
%NGC~4449 and the SMC because of its globular cluster habitat, a detonated
%C/O white dwarf could provide the copious amount of oxygen and velocity
%structure seen in RZ2109.
%Alternatively, Merritt, Schnittman, \& Komossa (2009) suggested
%that RZ2109 could be a star cluster ripped out of a galactic nucleus by a
%recoiling supermassice black hole.
The wide [O III] line might suggest a different scenario: an IMBH in RZ2109
has tidally-detonated a WD that passed sufficiently close in a
supernova-like event, as described in the numerical simulations of
Rosswog et al.\ (2008), and we are seeing the aftermath of this event,
as a supernova remnant is being formed. Fall-back of debris from the detonation
event could lead to the X-ray emission. If oxygen had
been torn from the WD during a previous passage, or if undetonated
oxygen in the outer layers of the WD was swept up by the fast-moving
expanding ejecta, such a wide [O III] line might result from the
photoionization of this material by the X-ray emission. The lack of an
appreciable ISM explains the lack of Balmer lines typically seen in Type Ia
supernova. Further work is needed to determine
if such a scenario can reproduce the observed X-ray and optical line emission
characteristics. Specifically, the physical conditions (density, temperature,
velocity) under which the shape
and strength of the [O III] line can be produced without accompanying
emission from Fe from the supernova ejecta will need to be investigated.
Such modeling will need to take into account the unusual conditions of such
a scenario compared to a classical Type Ia supernova, for example, the
potential presence of circumstellar oxygen that was stripped from the white
dwarf on previous encounters with the IMBH prior to the detonation event.
The long-term X-ray light curve variations of RZ2109 (Shih et al.\ 2008) will
also need to be shown to be consistent with a fallback scenario.

\acknowledgments

We thank Jon Miller, Tom Richtler, Stephanie Komossa, and Jari Kajava for
useful discussions. We also thank an anonymous referee for helpful suggestions.
This work was supported by NASA LTSA grant NNG05GE48G and {\it Chandra} grant
GO8-9087X.

\newpage
                                                                                
\begin{figure}
\plottwo{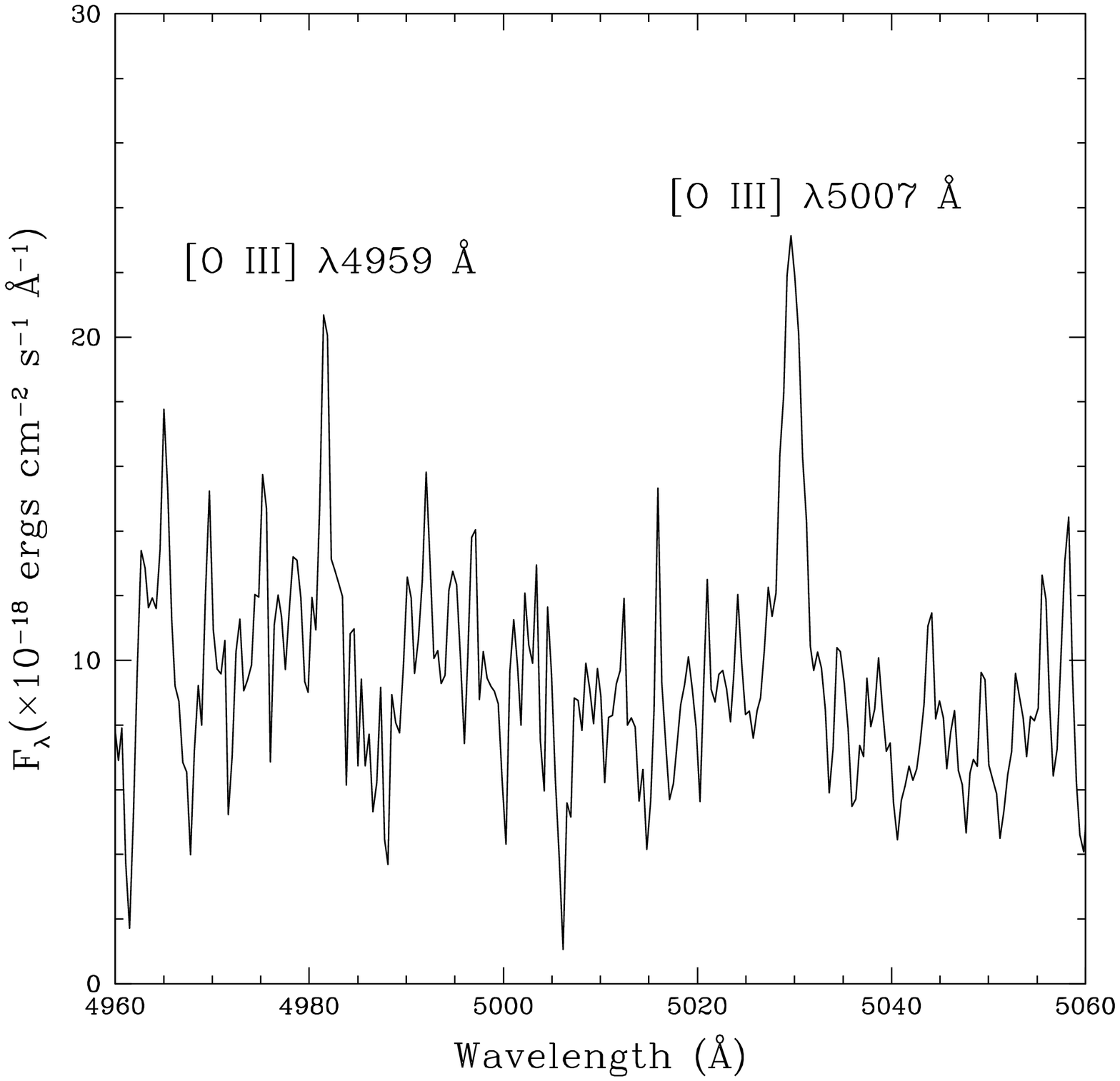}{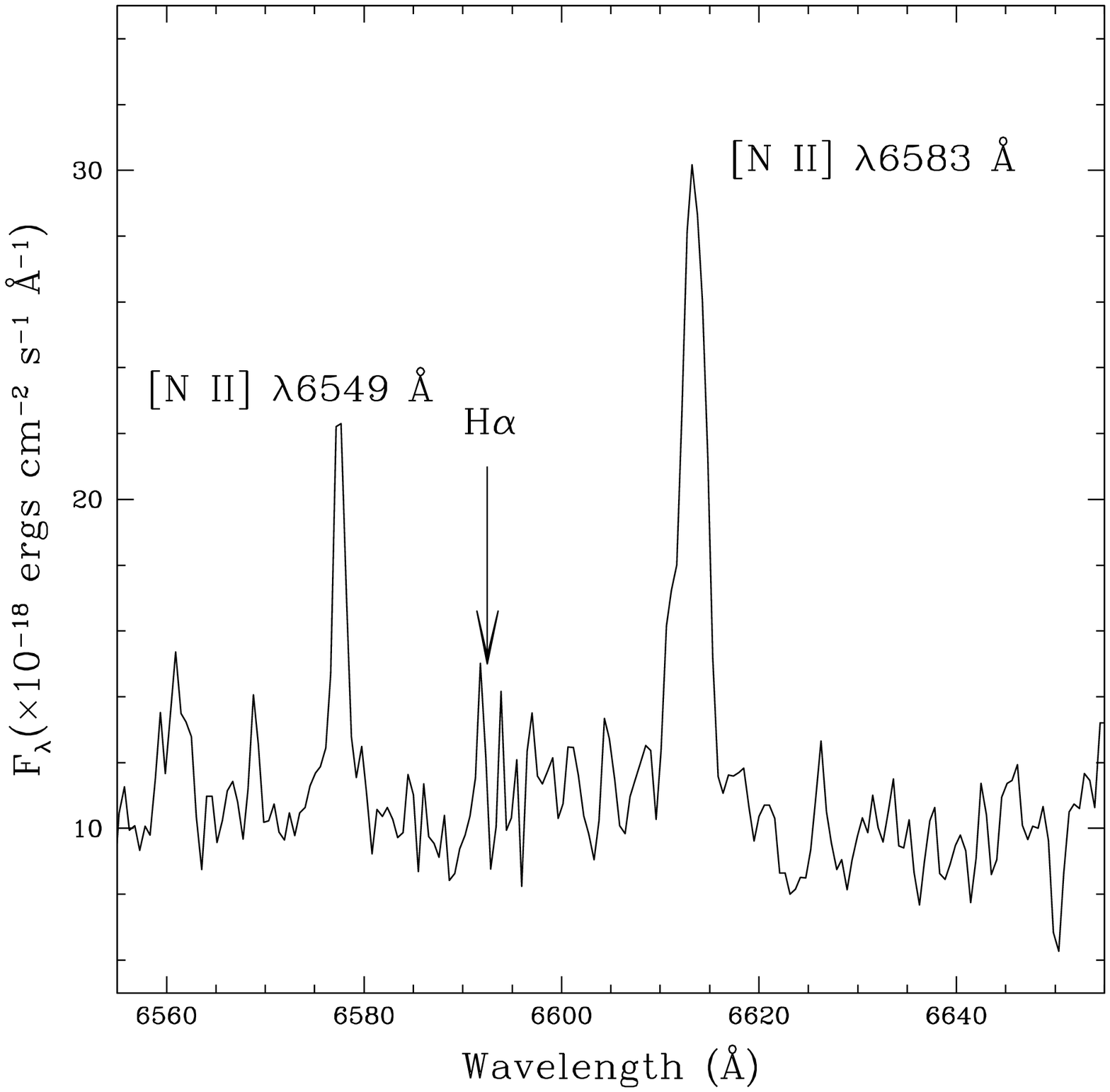}
\caption{{7.5 hr flux-calibrated MagE spectrum of the globular cluster harboring
CXOJ033831.8-352604 illustrating the [O III] ({\it left}) and [N II]
({\it right}) emission lines. Note the lack of H${\alpha}$.}
\label{fig:1}}
\end{figure}

\end{document}